\documentclass[a4paper,11pt]{article}
\usepackage{graphicx}
\usepackage{amssymb}
\usepackage{amsthm}
\usepackage{amsmath}
\usepackage{url}
\usepackage{subfigure}
\usepackage{color}
\usepackage{natbib}
\usepackage{url}
\usepackage{vmargin}
\setmarginsrb{2cm}{1.5cm}{2cm}{1.5cm}{0mm}{0mm}{0mm}{10mm}
\usepackage{titling}
\usepackage[breaklinks=true]{hyperref}
\usepackage{breakcites}

\begin{document}
\title{\vspace*{-1cm} \bfseries Qualitative comparisons of experimental results on deterministic freak wave generation based on modulational instability}
\author{\normalsize N. Karjanto$^\textmd{a,}$\thanks{\noindent Corresponding author. Tel.: +60 3 8924 8362; fax: +60 3 8924 2017. 
{\slshape E-mail address:} \texttt{natanael.karjanto@nottingham.edu.my} \newline
\copyright \ 2009 International Association for Hydro-environment Engineering and Research, Asia Pacific Division. Published by Elsevier~B.~V. All right reserved.
Journal of Hydro-environment Research {\bf 3} (2010) 186--192 \hfill doi:10.1016/j.jher.2009.10.008} \ and E. van Groesen$^\textmd{b,c}$\\
$^\textmd{a}${\small Department of Applied Mathematics, The University of Nottingham Malaysia Campus}\\
{\small Semenyih 43500, Selangor, Malaysia}\\
$^\textmd{b}${\small Department of Applied Mathematics, University of Twente, Enschede, The Netherlands}\\
$^\textmd{c}${\small LabMath-Indonesia, Bandung, Indonesia}}
\date{\footnotesize Received 27 May 2009; revised 9 October 2009; accepted 9 October 2009}
\maketitle

\begin{abstract}
A number of qualitative comparisons of experimental results on unidirectional freak wave generation in a hydrodynamic laboratory are presented in this paper. A nonlinear dispersive type of wave equation, the nonlinear Schr\"{o}dinger equation, is chosen as the theoretical model. A family of exact solutions of this equation the so-called Soliton on Finite Background describing modulational instability phenomenon is implemented in the experiments. It is observed that all experimental results show an amplitude increase according to the phenomenon. Both the carrier wave frequency and the modulation period are preserved during the wave propagation. As predicted by the theoretical model, a phase singularity is also observed in the experiments. Due to frequency downshift phenomenon, the experimental signal and spectrum lose their symmetric property. Another qualitative comparison indicates that the Wessel curves for the experimental results are the perturbed version of the theoretical ones.\\

\noindent
{\slshape Keywords:} Freak wave generation; Nonlinear Schr\"{o}dinger equation; Soliton on Finite Background; Modulational instability; Phase singularity;
Frequency downshift; Wessel curve.
\end{abstract}



\section{Introduction} \label{intro}
Freak wave events in the open oceans have attracted many scientists and engineers to study this topic and to predict the occurrence of the events.
A number of research directions have been conducted in the past decades, two of them are the deterministic and the statistical approaches. Several references on a statistical approach to study freak wave events are among others given by~\cite{Onorato01}, who study freak wave events in random oceanic sea state and~\cite{Janssen03} who studies freak waves and nonlinear four-wave interactions in the context of the stochastic approach. References on deterministic approach to study freak wave events are among others given by~\cite{Osborne00} and~\cite{Osborne01}, who study exact solutions of the nonlinear Schr\"{o}dinger (NLS) equation and~\cite{Henderson99}, who study computationally a fully nonlinear water wave equation and made a comparison with an exact solution of the NLS equation.

Our aim is to generate freak wave events in the wave basin of a hydrodynamic laboratory. We are interested in the unidirectional wave propagation and use the deterministic approach based on the theoretical model of the NLS equation. This equation describes a nonlinear dispersive type of surface wave envelopes and has applications not only in water waves but also in nonlinear optics, plasma physics and other applied fields. In particular, we select a family of exact solutions of the NLS equation that describe the Benjamin-Feir modulational instability phenomenon~\cite{Benjamin67}. This family of exact solution is known as the Soliton on Finite Background (SFB). Further, we implement the corresponding SFB wave signal to hydraulic wavemakers at the wave basin.

A set of experiments on freak wave generation has been conducted at the high-speed basin of the Maritime Research Institute Netherlands (MARIN) during summer 2004. The basin has a dimension of 200~m long, 4~m wide and the water depth of 3.55~m. Unidirectional waves are generated by a flap-type wavemaker at one side of the basin and the generated waves are absorbed by an artificial beach at the other side of the basin. The wavemaker hinge is located at 1.27~m above the basin floor. To measure the wave signal in the vertical direction, fourteen electronic wave gauges are installed along the basin. Two gauges are placed at 10~m and 11~m from the wavemaker, two at 40~m and 41~m, six around 100~m, two at 150~m and 151~m and the final two at 160~m and 161~m. More gauges were installed around 100~m than at other locations since the initial design of the experiments is to capture freak wave events around this position. However, our result shows that freak wave events occur at around 150~m and 160~m from the wavemaker.

Several aspects of these experimental results have been reported earlier. For instance, \cite{Huijsmans05} present a comparison between experimental results and numerical computations based on a fully nonlinear model for water waves, {\scshape hubris}. Furthermore, \cite{Karjanto06} discusses qualitative and quantitative comparisons between the experimental results and the theoretical model based on the SFB, including some remarks on variation in the model parameters, particularly the nonlinear coefficient of the NLS equation. This paper adds additional remarks on frequency downshift of the spectral peak in the frequency domain, as also confirmed experimentally by~\cite{Lake77}.

This paper is organized as follows. Section 2 discusses the theoretical model for freak wave based on the NLS equation. Section 3 discusses experimental results on freak wave generation, including wave signal, amplitude spectrum and Wessel curve comparisons. Finally, Section 4 gives conclusions and remarks.

\section{Theoretical model} \label{model}
Consider a Korteweg-de Vries (KdV) type of equation with exact dispersion relation~\citep{vanGroesen98}:
\begin{equation}
  \partial_{t}\eta + i \Omega(-i\partial_{x})\eta + \alpha\,\eta \partial_{x}\eta = 0, \qquad \alpha \in \mathbb{R}.
  \label{KdVexactdispersion}
\end{equation}
It is important to note that for our theoretical analysis, the variables are in normalized quantities but for the experiments, the variables are presented with units and are denoted with the subscript `lab'. Therefore, we have the relation $\eta = \eta_\textmd{lab}/h$, $x = x_\textmd{lab}/h$ and $t = t_\textmd{lab}\sqrt{g/h}$, where $h$ is the water depth and $g$ is the gravitational acceleration. Here $\eta(x,t)$ is the surface elevation from an equilibrium position. In this paper we consider physical wave fields in the $xt$-plane that are of the form of a wave packet $\eta$ as follows:
\begin{equation}
  \eta(x,t) = \epsilon A(\xi,\tau) e^{i(k_{0}x - \omega_{0}t)} + \textmd{higher order terms} + \textmd{complex conjugate}, \label{eta}
\end{equation}
where the wavenumber $k_{0}$ and the carrier wave frequency $\omega_{0}$ are related by the linear dispersion relation $\Omega(k) = \sqrt{k \tanh k}$. Note that these quantities are dimensionless. In laboratory variables, the linear dispersion relation is given by $\Omega_{\textmd{lab}}(k_{\textmd{lab}}) = \sqrt{g h \tanh (k_\textmd{lab}h)}$. After applying the multiple scale method with $\xi = \epsilon^{2}x$ and $\tau = \epsilon(t - x/\Omega'(k_{0}))$, the spatial NLS equation describes the approximate evolution equation of the complex-valued amplitude $A$ as follows:
\begin{equation}
\partial_{\xi}A + i \beta \partial_{\tau}^{2} A + i\gamma |A|^{2}A = 0, \qquad \beta \gamma > 0,
\label{spatialNLS}
\end{equation}
where $\beta$ and $\gamma$, both dependent on $\omega_{0}$, are the dispersive and the nonlinear coefficients, respectively.

In the following, the focusing case NLS equation has a number of exact solutions and particularly we are interested in a family of solutions describing freak wave model. This exact solution is a nonlinear extension of the Benjamin-Feir modulational instability. It is known as the Soliton on Finite Background (SFB), and can be explicitly expressed as follows:
\begin{equation}
  A(\xi,\tau) = A_{0}(\xi) \left(G(\xi,\tau)e^{i\phi(\xi)} - 1 \right), \label{SFB}
\end{equation}
where
\begin{equation}
  {\small
  G e^{i\phi} = \frac{\tilde{\nu}^{2} \cosh \sigma (\xi - \xi_{0}) - i \tilde{\sigma} \sinh \sigma (\xi - \xi_{0})}{\cosh \sigma (\xi - \xi_{0}) - \sqrt{1 - \frac{1}{2} \tilde{\nu}^{2}} \cos \nu (\tau - \tau_{0})}.
  }
\end{equation}
In this context, $\xi_{0}$ denotes the position and $\tau_{0} + 2n\pi/\nu$, for $n = 0, \pm 1, \pm 2, \dots$ denotes the time for which the SFB reaches its maxima. The quantity $\nu = r_{0} \sqrt{\gamma/\beta} \tilde{\nu}$, $0 < \tilde{\nu} < \sqrt{2}$ is the modulation frequency, $\sigma = \gamma r_{0}^{2} \tilde{\sigma}$, $\tilde{\sigma} = \tilde{\nu} \sqrt{2 - \tilde{\nu}^{2}}$ is the growth rate of the Benjamin-Feir instability and $A_{0} = r_{0} e^{-i\gamma r_{0}^{2} \xi}$ is the plane-wave solution of the NLS equation. This solution is discovered by~\cite{Akhmediev87}, see also~\citep{Akhmediev97}, and later independently using a different method by~\cite{Ablowitz90}. A number of authors has proposed the SFB as a model for freak wave events~\citep{Osborne00,Osborne01,Karjanto06,Dysthe99,Andonowati07}.

In the following section, we will discuss several qualitative comparisons between the experimental results and the theoretical model discussed in this section.

\section{Experimental results} \label{result}

A number of experiments have been conducted and the best result is selected and presented in this paper, i.e. the MARIN experiment No. 101062. The variation in the experiments depends on the corresponding theoretical values of the maximum amplitude and the modulation frequency. Although some waves were breaking before the wave signals reach the predicted extreme position at 150~m from the wavemaker, it is observed that all experimental signals show an amplitude increase as they propagate downstream. Interestingly, this amplitude increase obeys the nonlinear extension of Benjamin-Feir modulational instability phenomenon as described by the SFB solution of the NLS equation.

\subsection{Wave signal comparison}
There are two types of parameters involved in the experiments, i.e. basic and design parameters. The basic parameters are based on the model equation of the NLS equation and the design parameters are based on the variation in the SFB solution.
\begin{figure}[h!]
  \begin{center}
  \includegraphics[width=0.63\textwidth]{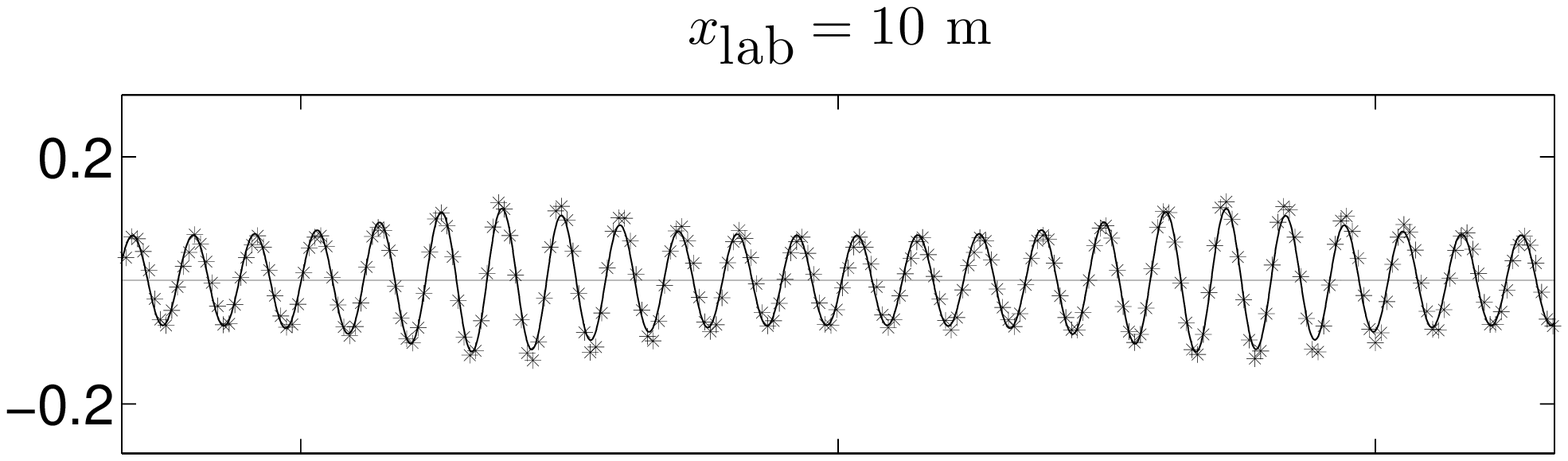}   \vspace*{0.2cm}\\
  \includegraphics[width=0.63\textwidth]{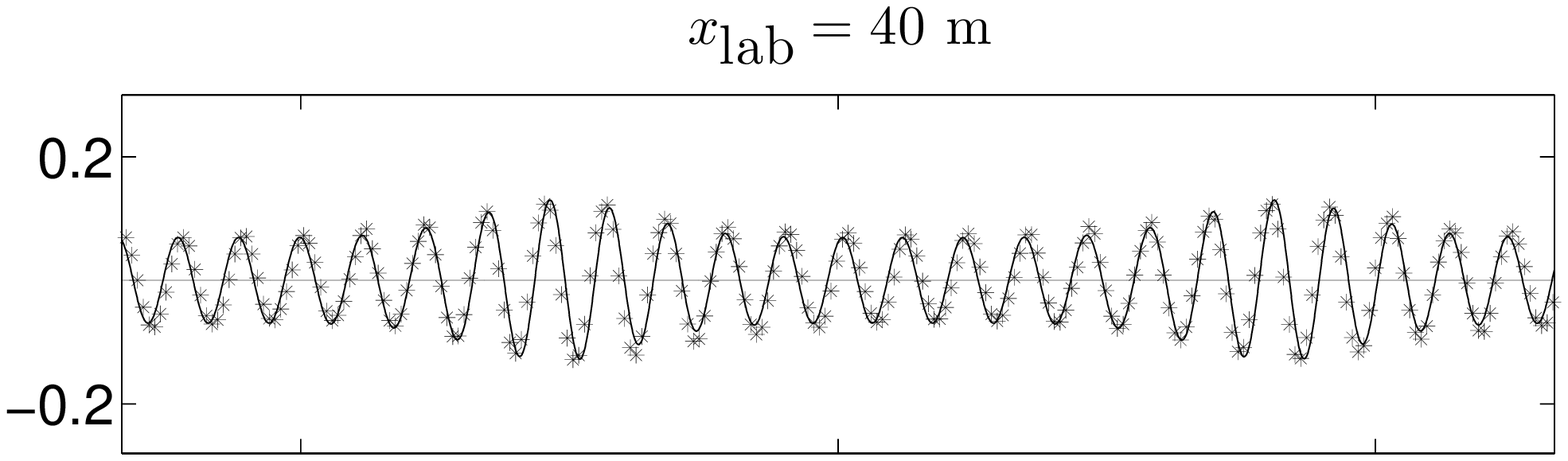}   \vspace*{0.2cm}\\
  \includegraphics[width=0.63\textwidth]{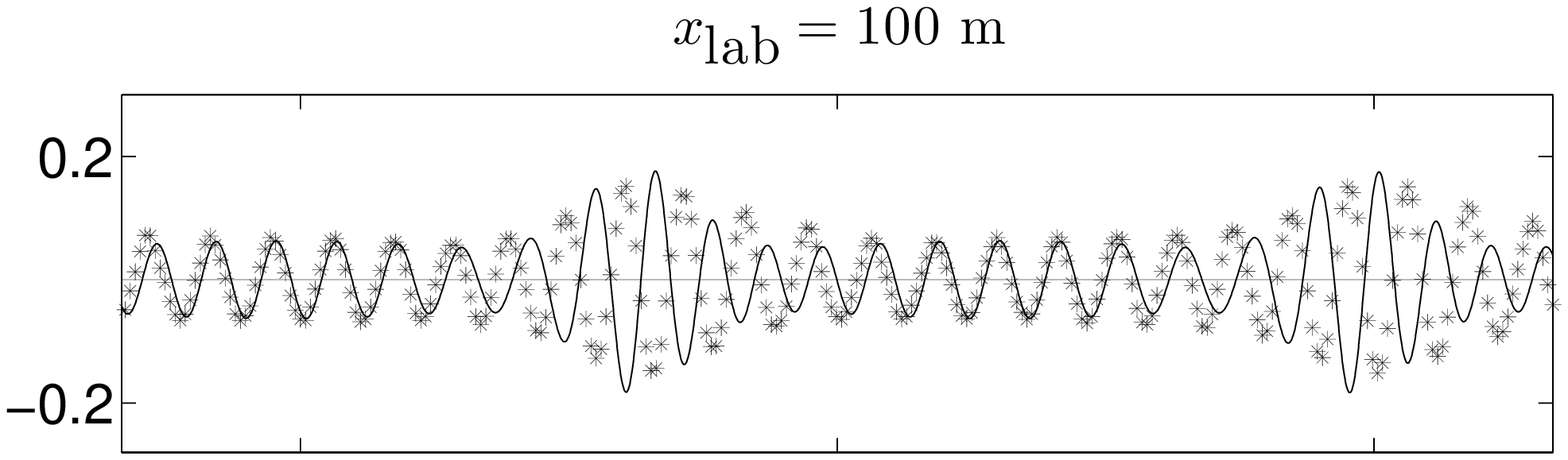}  \vspace*{0.2cm}\\
  \includegraphics[width=0.63\textwidth]{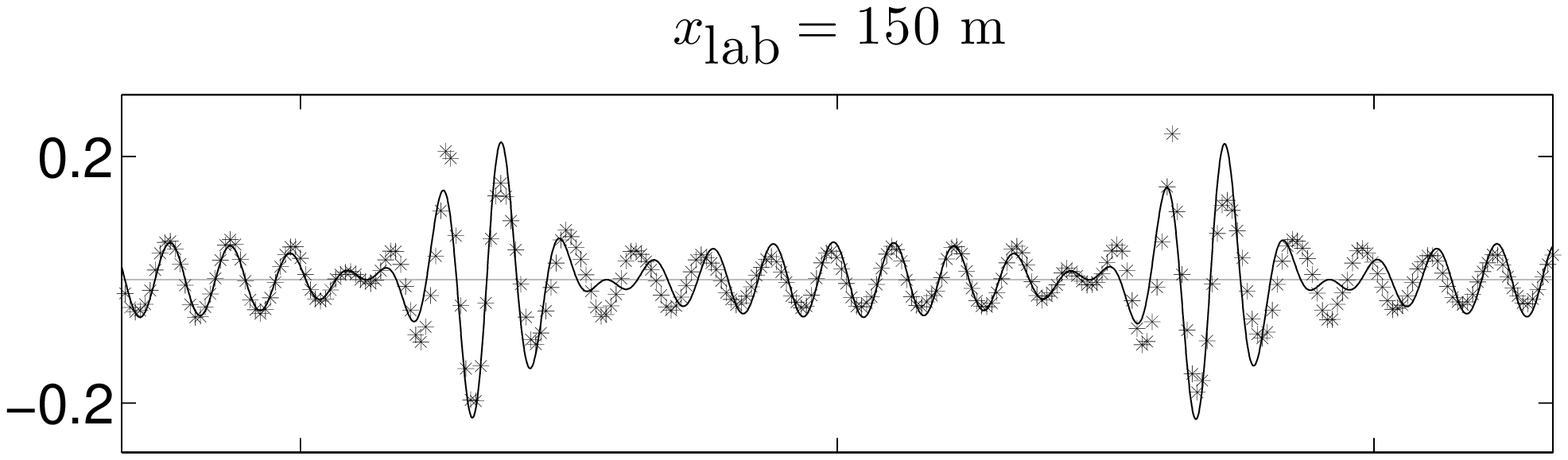}  \vspace*{0.2cm}\\
  \includegraphics[width=0.63\textwidth]{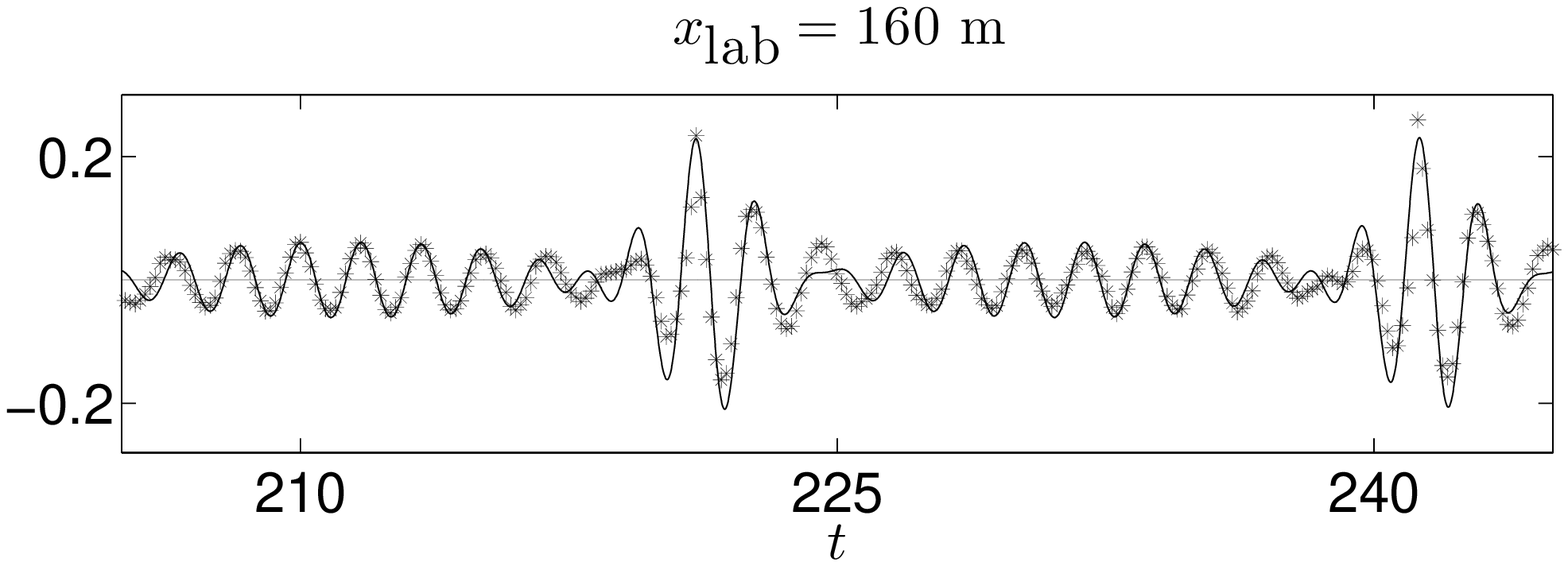}
  \caption[]{Plots of comparison between the theoretical SFB signal (solid) and the experimental signal (dotted) at 10~m, 40~m, 100~m, 150~m and 160~m from the wavemaker (relevant parameter values are given in the text). Some adjustments have been implemented by including the group velocity in a moving frame of reference so that the best match is obtained between theoretical and experimental signals.} \label{comparesignal}
  \end{center}
\end{figure}

Basic parameter values for the experiments are given as follows. The carrier wave frequency $\omega_{0\textmd{lab}} = 3.7284$~rad/s = 0.5934~Hz, the carrier wave period is 1.6852~s, the carrier wavenumber $k_{0\textmd{lab}} = 1.42$~m$^{-1}$, the carrier wavelength $\lambda_{0\textmd{lab}} = 4.43$~m, the predicted extreme position $x_\textmd{lab} = 150$~m, the dispersive coefficient $\beta = 1.01$ and the nonlinear coefficient $\gamma = 230.25$. The design parameters for a particular experiment presented in this paper are given as follows. The predicted maximum amplitude is 21.3~cm, the normalized modulation frequency is $\tilde{\nu} = 1$, which means that the modulation frequency is 0.3114~rad/sec = 0.0496~Hz, the modulation period is 20.18~sec, the number of waves in one modulation period is 11.98 and the initial steepness is 0.125.

A comparison between the theoretical signal based on the SFB and the experimental wave signal is displayed in Figure~\ref{comparesignal}. The comparison is given at $x_\textmd{lab} = 10$~m, where it is relatively close to the wavemaker, $x_\textmd{lab} = 40$~m, $x_\textmd{lab} = 100$~m, at $x_\textmd{lab} = 150$~m, where the extreme wave signal is expected to occur and $x_\textmd{lab} = 160$~m. It is observed that both theoretical and experimental wave signals show a good agreement when they are relatively close to the wavemaker. However, as they travel downstream, further away from the wavemaker, a difference starts to appear between the two wave signals. At 100~m from the wavemaker, we observe that a discrepancy starts to occur in particular at the large amplitude parts. A significant difference between the theory and the experiment is on the symmetry property. The theoretical signals maintain a symmetric structure within one modulation period. On the other hand, the experimental signals lose their symmetric property, particularly after they travel sufficiently far from the wavemaker. Rather, they show a deformed wave packet structure with a steep front and a flat rear within one modulation period.

The theoretical SFB signal shows phase singularity at the extreme position, a phenomenon closely related to wavefront dislocation~\citep{Karjanto06,Karjanto07}. This phenomenon occurs when the wave envelope vanishes, i.e. the complex-valued amplitude crosses the origin in the complex plane and therefore the phase is undefined and becomes singular. We can also observe in Figure~\ref{comparesignal} that at 150~m, there is a sudden jump in the distance between the crests, indicating that the change in local frequency is very large. The observance of the wave envelope vanishing and the large change in local frequency are indications the occurrence of phase singularity for both theoretical and experimental signals. For the theoretical SFB signal, there is one pair of phase singularities in one modulation period. In the experimental signal, however, only one singularity occurs in one modulation period. See Figure~\ref{comparesignal} particularly at 150~m and 160~m from the wavemaker. The difference in the occurrence of phase singularity is related to the asymmetric structure of the wave signal mentioned above as well as the frequency downshift phenomenon that will be discussed in the following subsection.

\subsection{Amplitude spectrum comparison}

Another interesting property of the wave signal can be observed in the frequency domain. By applying the Fourier transform to the wave signal, we obtain the corresponding wave spectrum. This spectrum is a complex-valued quantity, depends on the frequency $\omega$ and the position from the wavemaker $x_\textmd{lab}$ and is denoted as $a(\omega)$. In this paper, we present the plots of the absolute amplitude spectrum $|a(\omega)|$ at different positions. By comparing the absolute amplitude spectrum evolution between the theoretical and the experimental wave signals, we observe that frequency downshift phenomenon is observed in the experimental spectrum and not in the theoretical one.
\begin{figure}[h!]
  \begin{center}
  \includegraphics[width=0.54\textwidth]{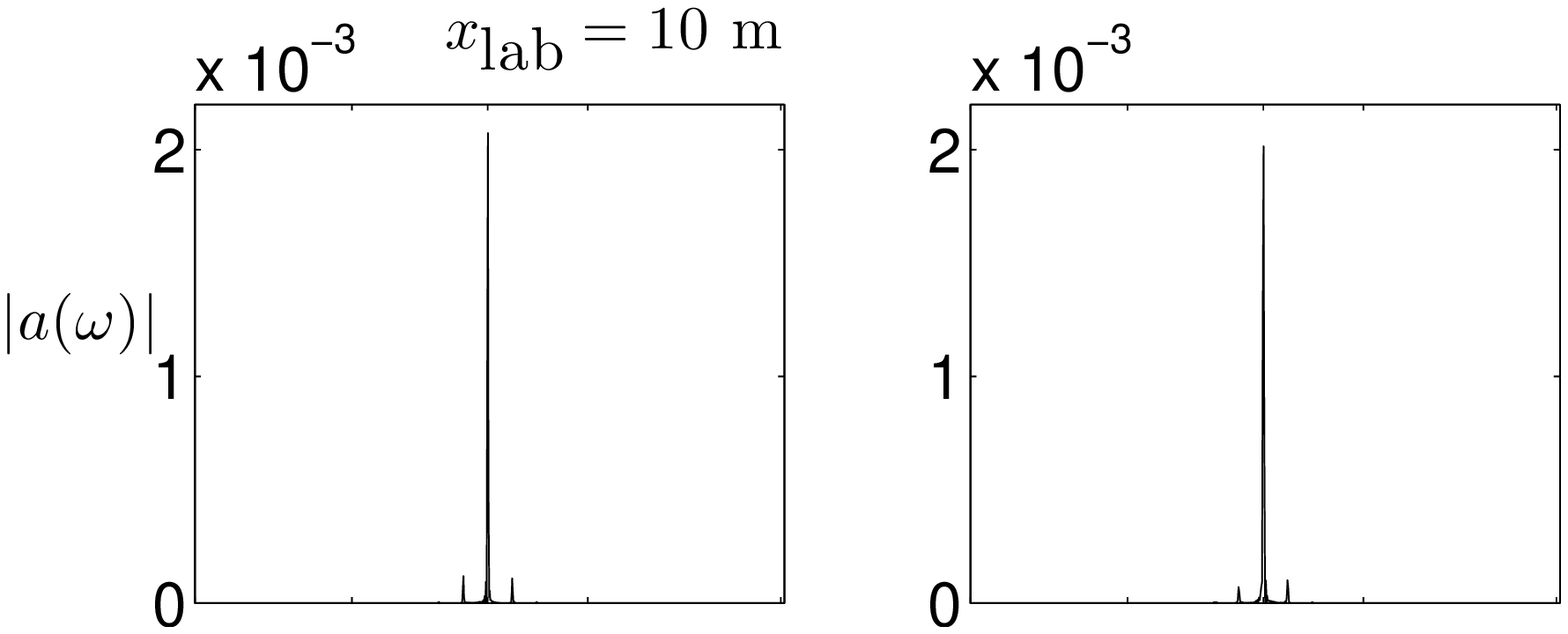}  \vspace*{0.12cm}\\ 
  \includegraphics[width=0.54\textwidth]{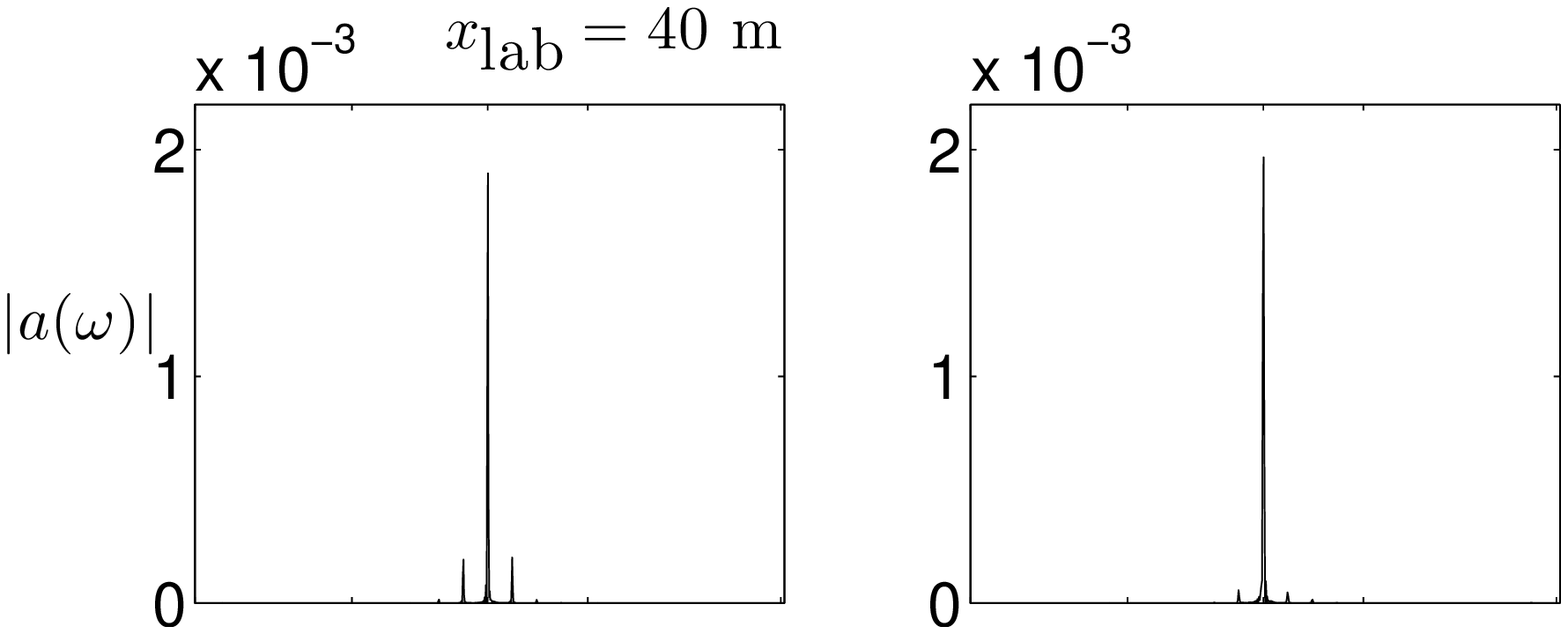}  \vspace*{0.12cm}\\
  \includegraphics[width=0.54\textwidth]{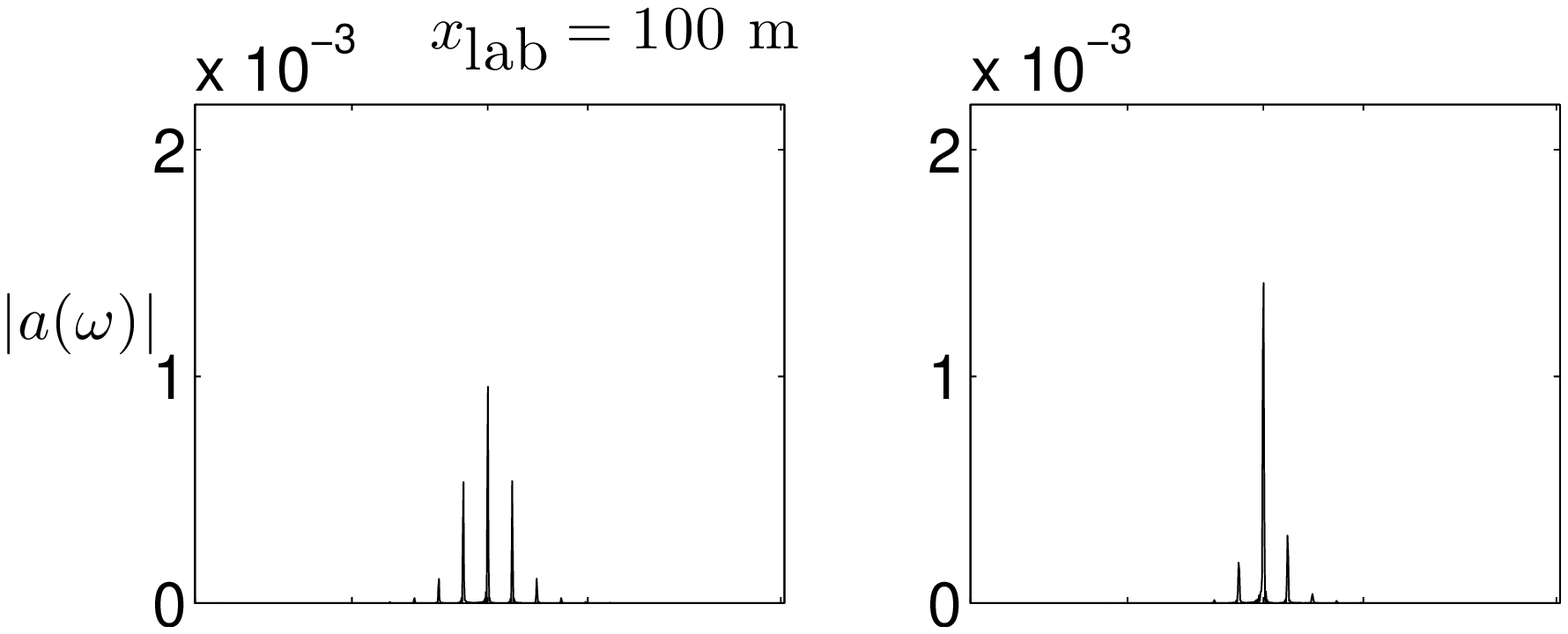} \vspace*{0.12cm}\\
  \includegraphics[width=0.54\textwidth]{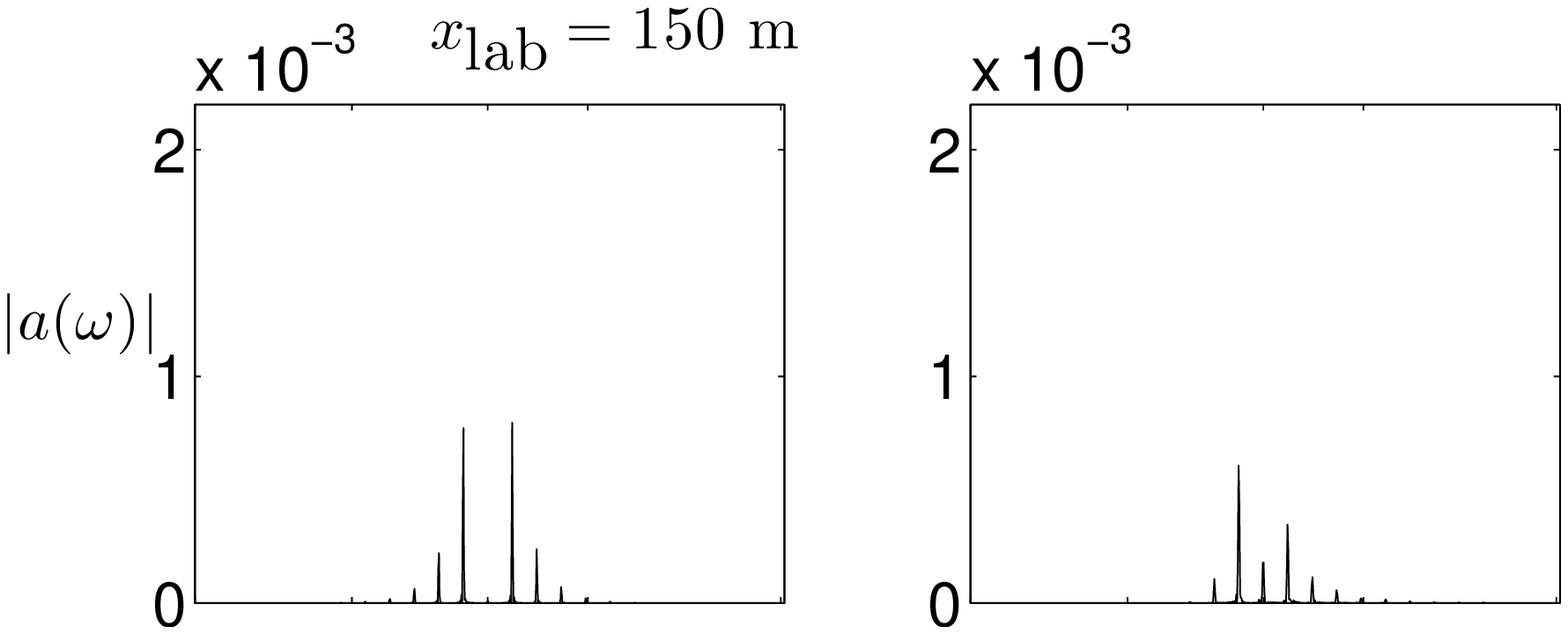} \vspace*{0.12cm}\\
  \includegraphics[width=0.54\textwidth]{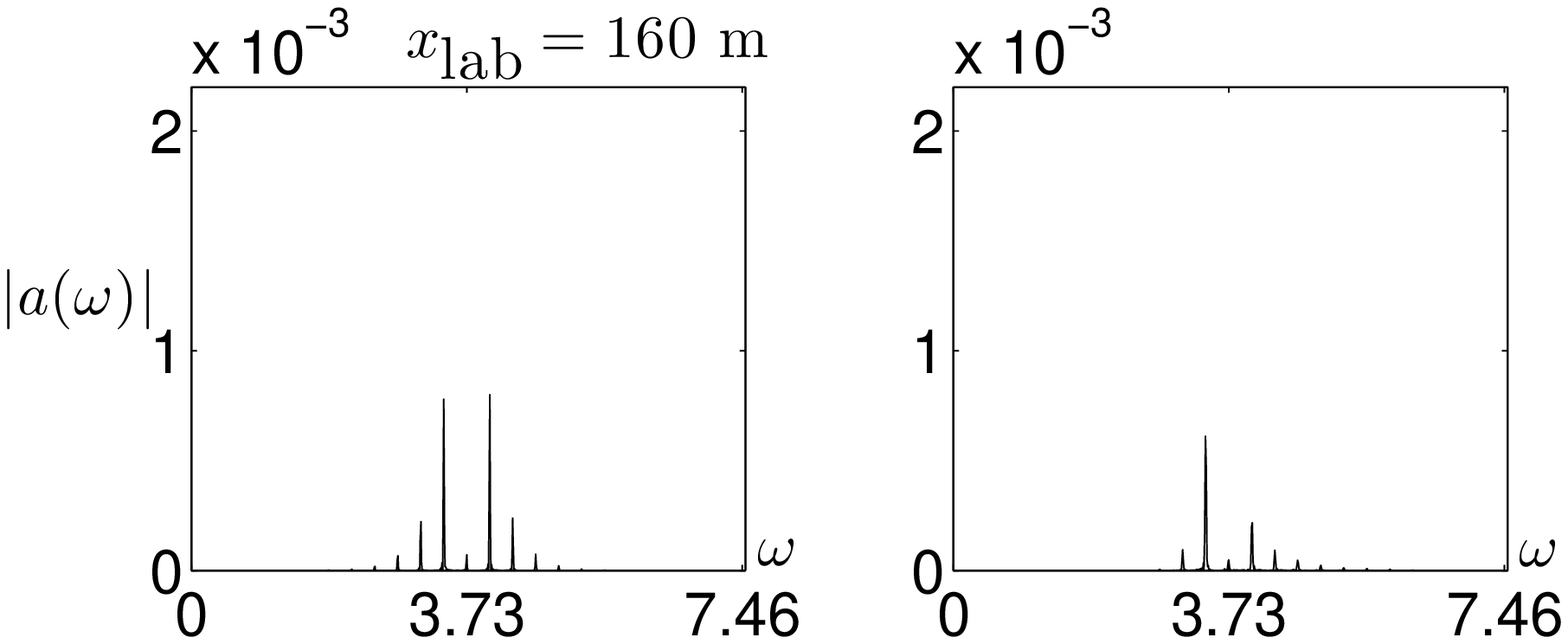}
  \caption[]{Plots of comparison between the theoretical SFB spectrum (left column) and the experimental spectrum (right column) at 10~m, 40~m, 100~m, 150~m and 160~m from the wavemaker. The axes are presented in laboratory variables, the frequency and the amplitude spectrum are given in rad/sec and in meter, respectively.} \label{comparespectrum}
  \end{center}
\end{figure}

Comparison plots of the theoretical SFB and the experimental absolute amplitude spectra at different positions are shown in Figure~\ref{comparespectrum}. We observe that at 10~m from the wavemaker, both theoretical and experimental spectra initially consist of the carrier wave frequency $\omega_{0}$ and one pair of sidebands $\omega \pm \nu$ components. As the wave travels downstream, the corresponding spectra also experience some changes. For the theoretical SFB spectrum, the magnitude of the carrier wave frequency component decreases, the magnitude of the first sideband frequency components increase and components of higher sideband frequencies appear in the spectrum domain. Each pair of the sideband frequency components grows with the same rate so that the spectrum remains symmetric with respect to the central component. After reaching the extreme position, the spectrum will return to the original form.

For the experimental spectrum, the magnitude of the carrier wave frequency component also decreases as the wave travels downstream. However, the lower sideband component grows at a faster rate than the upper sideband component. As a consequence, the peak of the spectrum shift to lower frequencies as the wave amplitude increases, known as `frequency downshift' phenomenon~\citep{Mei05}. Due to this frequency downshift, the spectrum loses its symmetric property, as is also the case for the wave signal. It has been widely believed that frequency downshift is attributable to the combined effects of nonlinear wave interactions and dissipation due to breaking, in which the latter effect is not really described by the NLS equation~\citep{Hara91}.

The frequency downshift phenomenon in the context of nonlinear wave interactions is observed for the first time by~\cite{Lake77} when they reported experimental results of a nonlinear wave train evolution on deep water. Some earlier references on frequency downshift phenomenon are found in the wind-wave interaction, among others are~\citep{Pierson71,Hasselmann73}. From the experiments conducted by~\cite{Melville82}, it is discovered that for rather steep waves, the onset of breaking corresponds to the onset of the asymmetric development of the sideband frequencies. In this case, the downshift is not reversible. A theoretical examination on frequency downshift in narrow banded surface waves under the influence of wind blowing in the direction of wave propagation is explained in~\citep{Hara91}. Two essential factors to cause downshift are sufficient nonlinearity to produce asymmetry in the spectral distribution and the nonlinear dissipation that affects the higher spectral components~\citep{Kato95}. However, \cite{Trulsen97} showed that dissipation may not be necessary to produce a permanent downshift in three-dimensional wave trains in deep water, see also~\citep{Trulsen99}.

In order to obtain a better agreement between theoretical and experimental results, several experts in the field have suggested improving the model equation for the wave packet evolution describing freak wave events. Fuller theories based either on Zakharov's integral equation~\citep{Yuen82} or Dysthe's fourth-order extension to the NLS equation, also known as the modified NLS equation~\citep{Dysthe79,Lo85,Trulsen96} can predict the occurrence of asymmetric wave signal~\citep{Hara91}. A numerical study based on the modified NLS equation and the comparison with the experiments have been reported by~\cite{Lo85}.

The difference in the occurrence of phase singularity between the experimental results and the theoretical SFB can also be comprehended by qualitatively comparing the Wessel curves in the complex plane, which will be discussed in the following subsection.

\subsection{Wessel curve comparison}

Recall again that the family of SFB solution~\eqref{SFB} is a complex-valued function and thus we may describe its evolution in the complex plane. In order to obtain a better visualization display, the plane-wave solution will be excluded from the SFB expression since it simply produces oscillation effect in the complex plane and thus overshadows the desired evolution curves. Hence, we will only consider the complex-valued function $F(\xi,\tau) := G(\xi,\tau) e^{i\phi(\xi)} - 1$.
\begin{figure}[h!]
  \begin{center}
    \subfigure[]{\includegraphics[width=0.29\textheight, bb = 36 211 537 612]{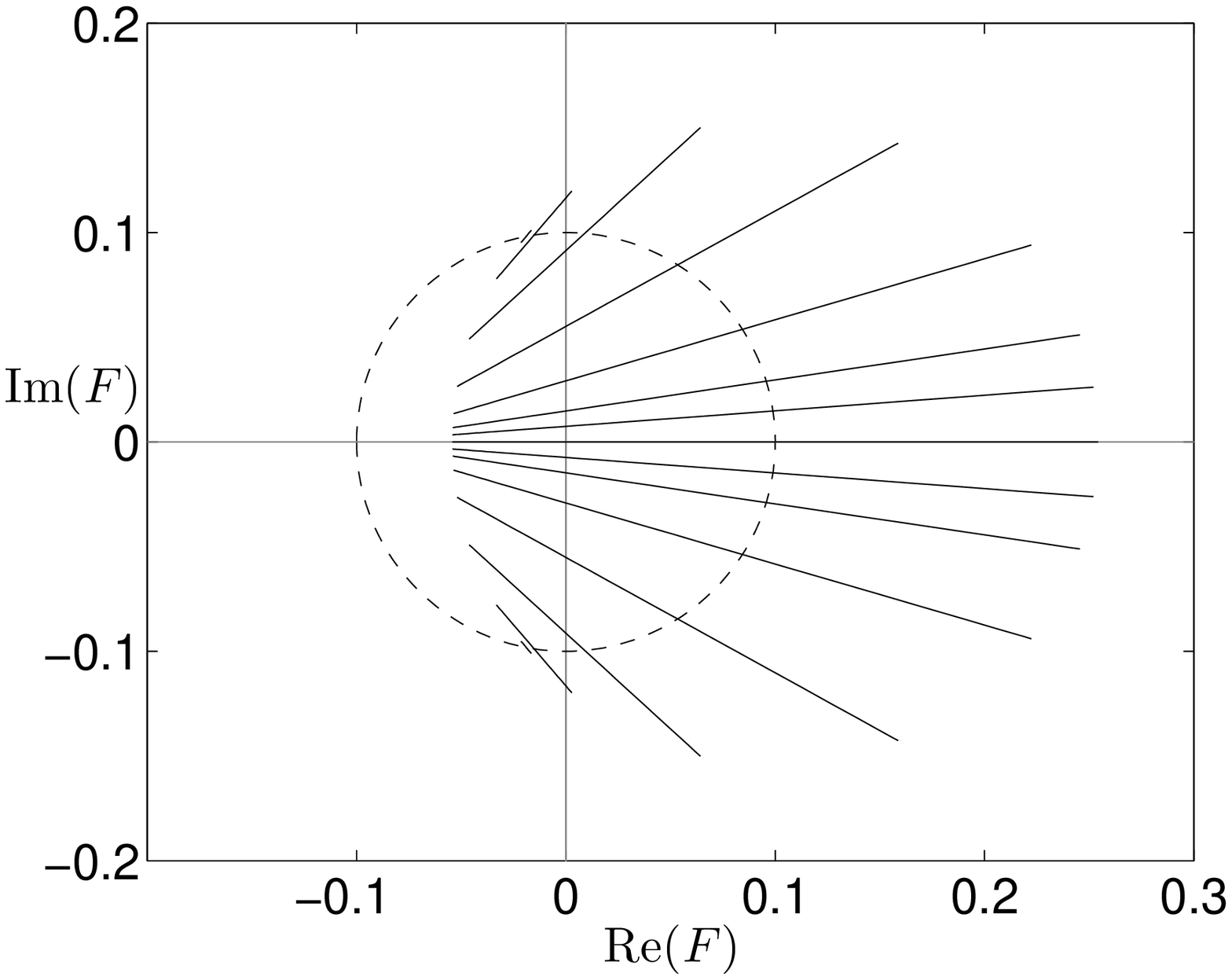}}    \hspace{0.5cm}
    \subfigure[]{\includegraphics[width=0.29\textheight]{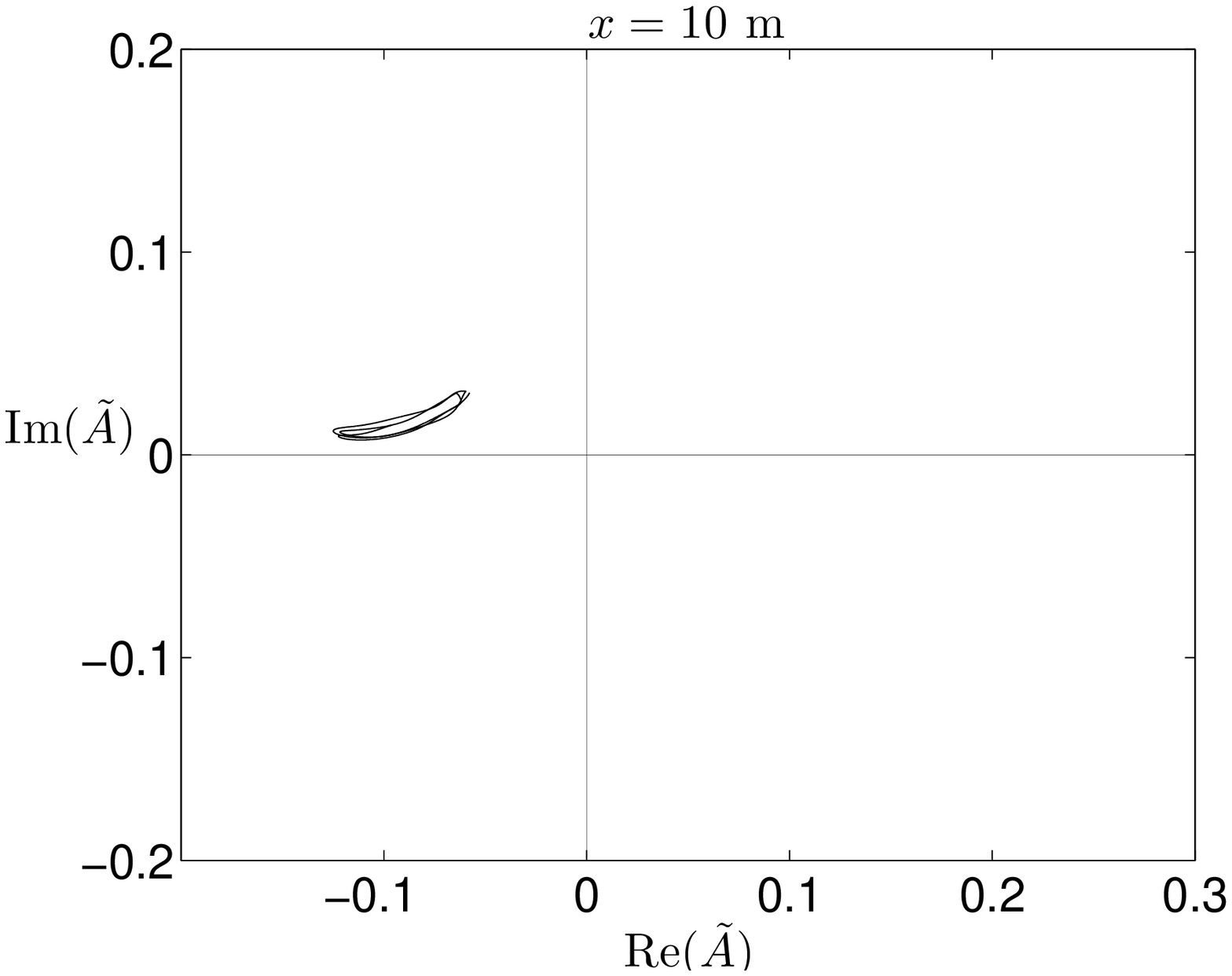}} \\ 
    \subfigure[]{\includegraphics[width=0.29\textheight]{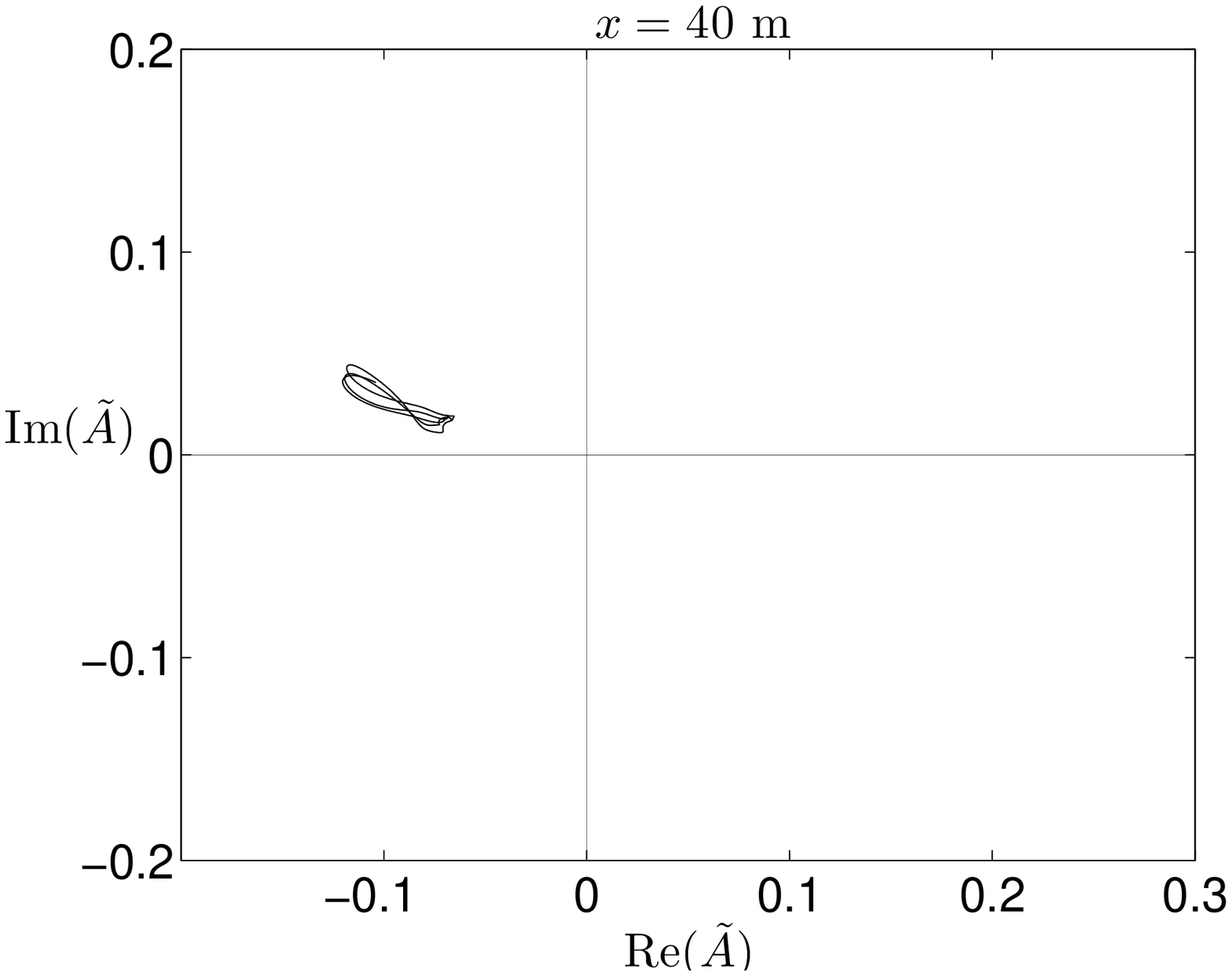}} \hspace{0.5cm}
    \subfigure[]{\includegraphics[width=0.29\textheight]{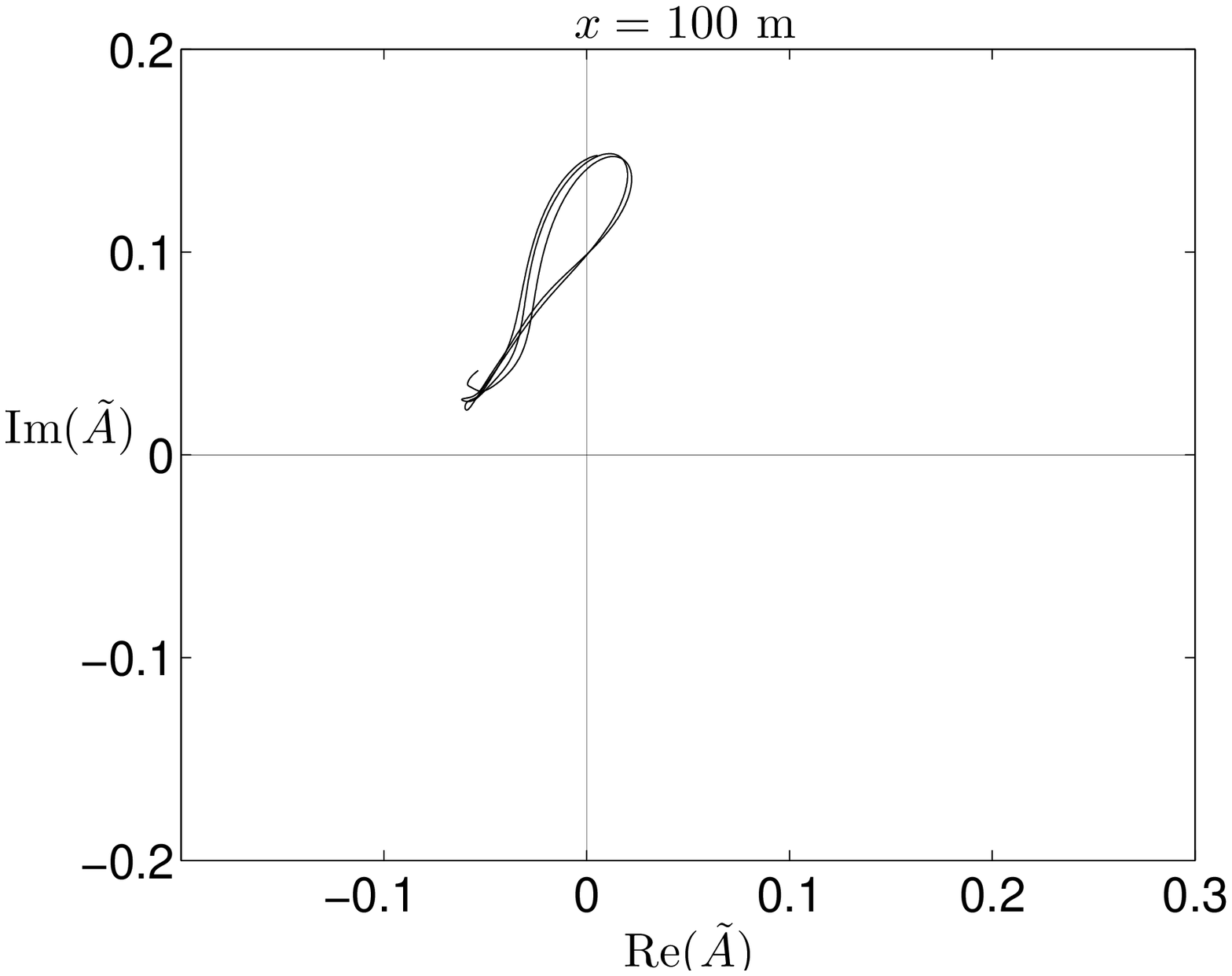}} \\ 
    \subfigure[]{\includegraphics[width=0.29\textheight]{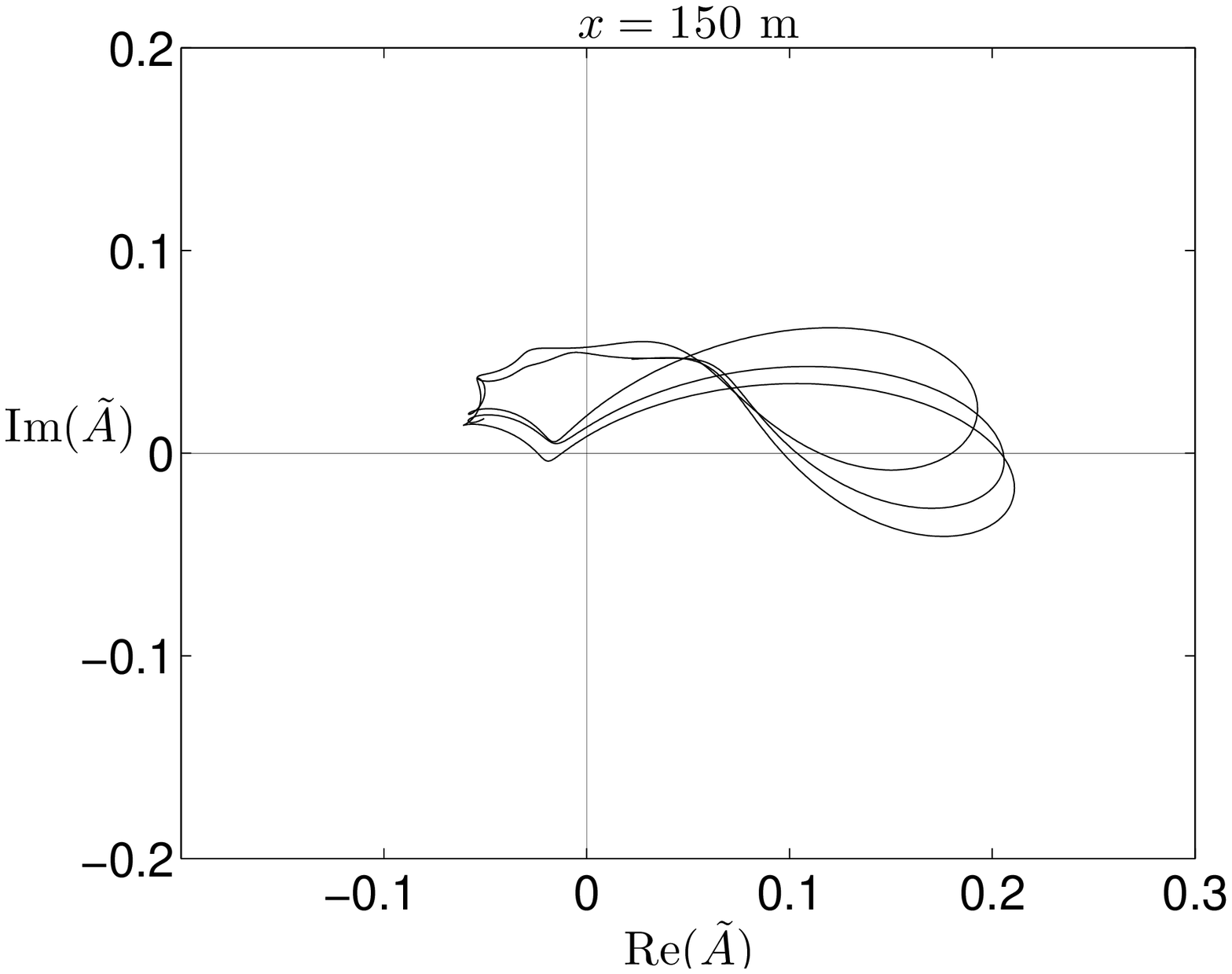}} \hspace{0.5cm}
    \subfigure[]{\includegraphics[width=0.29\textheight]{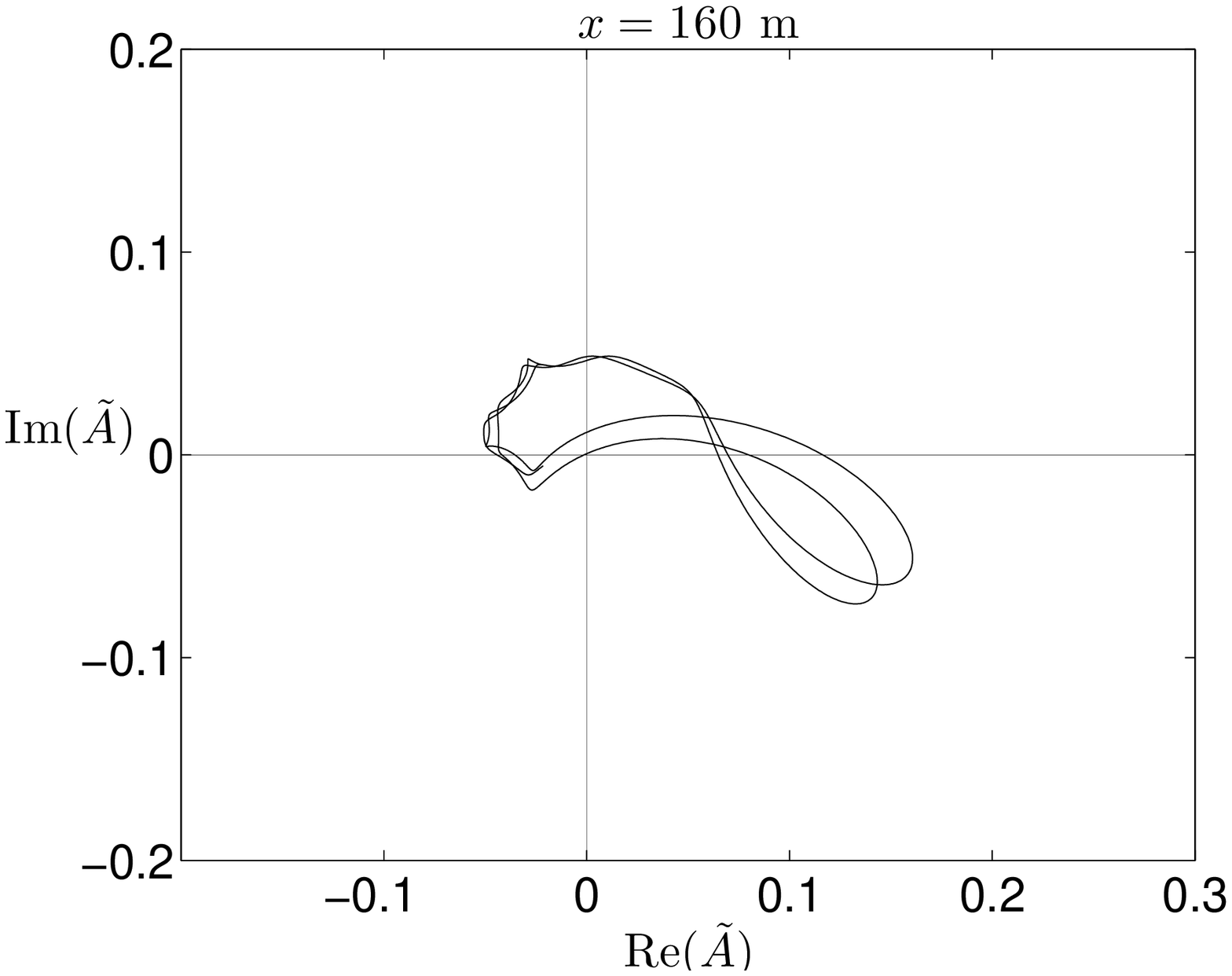}}
    \caption{Plots of the spatial Wessel curve of (a) the theoretical SFB and (b)--(f) the experimental signal at different positions: $x_{\textmd{lab}} = 10$~m, 40~m, 100~m, 150~m and 160~m. Both $F$ and $\tilde{A}$ correspond to complex-valued amplitudes for the theoretical and experimental signals, respectively, after removing the oscillating components.}
    \label{Wesselcurve}
  \end{center}
\end{figure}

This type of dynamical evolution formerly is known as `Argand diagram representation', where the horizontal and vertical axes correspond to the real and the imaginary parts of $F$, respectively~\citep{Karjanto06}. In this paper, we now refer this representation by introducing a novel term called `Wessel curve'. One compelling reason to do so is in honoring a Norwegian surveyor Caspar Wessel (1745-1818), who gave the first geometrical interpretation of complex number in 1799, earlier than a French mathematician Jean-Robert Argand (1768-1822)~\citep{encyclopedia,Argand}. Therefore, the well-known Argand diagram is actually the Wessel diagram.

There are two types of Wessel curve: spatial and temporal Wessel curves. The spatial Wessel curve refers to the evolution of the SFB parameterized in time $\tau$ at several positions $\xi$. The temporal Wessel curve refers to the evolution of the SFB parameterized in position $\xi$, for different time frames~$\tau$. In this paper, we only compare the spatial Wessel curve between the theoretical SFB and the experimental results. Although the temporal Wessel curve has interesting characteristics too, we are not able to compare with the experimental results since the number of our measuring positions are very few to produce presentable curves. The readers who are interested in the temporal Wessel curve for the theoretical SFB are encouraged to consult~\citep{Akhmediev97}.

The spatial Wessel curve for the theoretical SFB is a set of straight lines centered at $(-1,0)$ in the complex plane. At a particular position, a portion of these lines is passed twice during one modulation period. Each line forms an angle with respect to the positive real axis and this angle corresponds to the displaced amplitude $\phi(\xi)$. The lines are positioned above the real axis for $\xi < 0$ and below it for $\xi > 0$. At $\xi = 0$, it lies precisely on the real axis. For normalized modulation frequency $0 < \tilde{\nu} < \sqrt{3/2}$, this particular line passes the origin twice during one modulation period. This corresponds to the extreme position $x_{\textmd{lab}}$ where freak waves occur in the wave tank. At this position, one pair of phase singularities is observed during one modulation period. The spatial Wessel curve for the theoretical SFB is depicted in Figure~\ref{Wesselcurve}(a).

Earlier in this section, we have pointed out that the experimental signals lose the symmetric structure as they travel downstream away from the wavemaker. After investigating the amplitude spectrum evolution in the frequency domain, we understand that the difference in the symmetry property is closely related to the frequency downshift phenomenon. Now in this subsection, we observe that the Wessel curve for the experimental results is a collection of twisted ellipse-like curves in the complex plane, as can be seen in Figure~\ref{Wesselcurve}(b)--~\ref{Wesselcurve}(f). Qualitatively, we interpret that this experimental Wessel curve is comparable to the theoretical one after imposing a small perturbation. This situation can be explained by keeping in mind that those straight lines are actually `flattened' closed curves with zero area around their boundary. Perturbations will deform the straight lines. It seems that the straight lines are `blown-up' and become bent, almost closed, curves with self-intersection as in Figure~\ref{Wesselcurve}(b)--~\ref{Wesselcurve}(f) that make a slow rotational motion. A more detailed perturbation analysis method would require an extensive mathematical study of the stability of the SFB, which has not yet been performed and is outside the scope of this contribution.

As a consequence, only one phase singularity is visible during one modulation period for the experimental signals, in comparison to one pair of phase singularities for the theoretical SFB signals. As can be observed in Figure~\ref{Wesselcurve}(e) and~\ref{Wesselcurve}(f), the curves cross through the origin only once during one modulation period. Actually, the particular curves that really cross the origin were not really captured at the limited positions of measurement. However, their positions which are very close to the origin are sufficiently good in displaying the phase singularity.

\section{Conclusion and remark}

After collecting experimental data and perform some analysis, we observe that all experimental results show amplitude increase according to the Benjamin-Feir modulational instability phenomenon corresponding to the SFB solution of the NLS equation. The carrier wave frequency and the modulation period are conserved during the wave propagation in the wave basin. A significant difference is that the experimental wave signal does not preserve the symmetry structure as possessed by the SFB wave signal. 

Nevertheless, the experimental signal shows phase singularity, a phenomenon closely related to wavefront dislocation. In one modulation period, the SFB wave signal has one pair of phase singularities but the experimental signal has only one singularity in this same time interval. In the frequency domain, frequency downshift phenomenon is observed in the evolution of the experimental spectrum but it is not in the theoretical SFB spectrum, where each pair of sideband frequency components grows and decays at the same rate. By qualitatively comparing the (spatial) Wessel curves between the experimental result and the theoretical SFB, we observed that the experimental Wessel curves can be interpreted as the perturbed version of the theoretical ones. Perturbations will deform the theoretical Wessel curves into bent and almost closed curves with self-intersection that make a slow rotational motion. This comparison also confirms the occurrence of only one phase singularity during one modulation period for the experimental results.

\section{Future research}

One possible research direction is by improving the theoretical model for describing the wave evolution in the wave tank. A more accurate modeling for unidirectional surface waves elevation has been proposed by~\cite{vanGroesen07}, known as the AB~equation. Using the Hamiltonian of surface waves, they derive higher order KdV equations for waves above finite depth. Furthermore, the accuracy of this new model has been tested in two different cases: a good agreement of wave profiles and wave speed with respect to Stokes' waves and a good agreement of propagation speeds and wave envelope distortions in the context of bichromatic wave groups~\citep{vanGroesen10}. Since the AB~equation shows good agreements in these two cases, it would be interesting to explore further in connection with freak wave generation for unidirectional waves. It would be interesting if a couple similar phenomena which are discussed in this article will be observed too. Even more, perhaps some other intriguing phenomena are still yet to be discovered.

\subsection*{Acknowledgement}
{\small
The research is supported by the project TWI.5374 of the Netherlands Organization of Scientific Research NWO, subdivision Applied Sciences STW. The authors would also like to acknowledge the Chair of Applied Analysis and Mathematical Physics at University of Twente, MARIN, J.M. Burgerscentrum, European-Union Jakarta (project SPF 2004/079-057), Faculty of Engineering at The University of Nottingham Malaysia Campus, Dr. Andonowati (Bandung Institute of Technology and LabMath-Indonesia), Prof. Ren\'{e} Huijsmans (TU Delft and MARIN), Gert Klopman (University of Twente and Albatros Flow Research), Dr. Miguel Onorato (University of Torino), Prof. Christian Kharif (University of Aix-Marseille) and Prof. Karsten Trulsen (University of Oslo).
\par}

\end{document}